# Design of an Efficient Neural Key Distribution Centre

MISS. SAHANA S.BISALAPUR
*M.Tech Computer Science and Engineering*
sahanabisalapur@gmail.com

ABSTRACT

*The goal of any cryptographic system is the exchange of information among the intended users without any leakage of information to others who may have unauthorized access to it. A common secret key could be created over a public channel accessible to any opponent. Neural networks can be used to generate common secret key. In case of neural cryptography, both the communicating networks receive an identical input vector, generate an output bit and are trained based on the output bit. The two networks and their weight vectors exhibit a novel phenomenon, where the networks synchronize to a state with identical time-dependent weights. The generated secret key over a public channel is used for encrypting and decrypting the information being sent on the channel. This secret key is distributed to the other vendor efficiently by using an agent based approach.*

1. INTRODUCTION

Artificial neural networks are parallel adaptive networks consisting of simple nonlinear computing elements called neurons which are intended to abstract and model some of the functionality of the human nervous system in an attempt to partially capture some of its computational strengths. Neural networks [1] are non-linear statistical data modeling tools. They can be used to model complex relationships between inputs and outputs or to find patterns in data. A phenomenon of neural network is applied in cryptography systems. This is used for generating secret key over public channel.

2. BACKGROUND

Cryptography is the practice and study of hiding information. It is an essential aspect for secure communication. Cryptography not only protects data from theft or alternation but also can be used for user authentication. Cryptography can also be defined as the conversion of data into a scrambled code that can be deciphered and sent across a public or private network. Cryptography [2] uses two main styles or forms of encrypting data; symmetrical and asymmetrical.

*2.1. Secret Key Cryptography*

In our work we are using symmetric key which use the same key for encryption as they do for decryption. With *secret key cryptography*, a single key is used for both encryption and decryption. As shown in the figure 1, the sender uses the key (or some set of rules) to encrypt the plaintext and sends the ciphertext to the receiver. The receiver applies the same key (or ruleset) to decrypt the message and recover the plaintext. Because a single key is used for both functions, secret key cryptography is also called *symmetric encryption*.

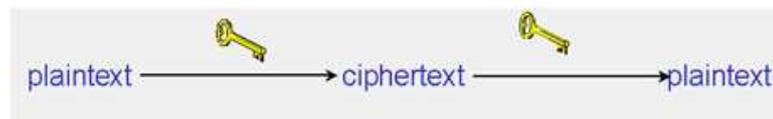

Figure 1: *Secret Key Cryptography*

*2.2. Secret Key Generation*

In cryptography pseudorandom number generators (PRNG's) were used to generate secret keys between two communicating parties. These typically start with a "seed" quantity and use numeric or logical operations to produce a sequence of values. A typical pseudo-random number generation technique is known as a linear congruence pseudorandom number generator. These are the mechanisms used by real-world secure systems to generate cryptographic keys, initialization vectors, \random" nonce's and other values assumed to be random. But here there are some possible attacks against PRNG's [3]. Here an attacker may cause a given PRNG to fail to appear random, or ways he can use knowledge of some PRNG outputs (such as initialization vectors) to guess other PRNG outputs (such as secret key). Hence to overcome this disadvantage neural network is used in cryptography to generate the secret key.

*2.3. Agent Approach*

The agents are autonomous i.e. they are capable of acting independently. An agent is anything that can be viewed as perceiving its environment through sensors and acting upon that environment through effectors [4]. Mobile agents are the basis of an emerging technology that promises to make it very much easier to design, implement, and maintain distributed systems [5]. We have found that mobile agents reduce network traffic, provide an effective means of overcoming network latency, and perhaps most importantly, through their ability to operate asynchronously and autonomously of the process that created them, help us to construct more robust and fault tolerant systems.

3. NEURAL CRYPTOGRAPHY

*3.1. Interacting Neural Network and Cryptography*

Two identical dynamical systems, starting from different initial conditions, can be synchronized by a common externalsignal which is coupled to the two systems. Two networks which are trained on their mutual output can synchronize to a time dependent state of identical synaptic weights [6]. This phenomenon is also applied to cryptography [7]. Neural networks learn from examples. This concept has extensively been investigated using models and methods of statistical mechanics [8] [9]. A"teacher" network is presenting input/output pairs of high dimensional data, and a"student" network is being trained on these data. Training means, that synaptic weights adopt by simple rules to the input/output pairs. After the training phase the student is able to generalize: It can classify – with some probability – an input pattern which did not belong to the training set. In this case, the two partners A and B do not have to share a common secret but use their identical weights as a secret key needed for encryption. In neural network an attacker E who knows all the details of the algorithm and records any communication transmitted through this channel finds it difficult to synchronize with the parties, and hence to calculate the common secret key. We assume that the attacker E knows the algorithm, the sequence of input vectors and the sequence of output bits. Initial weight vectors and calculate the ones which are consistent with the input/output sequence. It has been shown, that all of these initial states move towards the same final weight vector, the key is unique [10]. However, this task is computationally infeasible. In principle, E could start from all of the Synchronization by mutual learning (A and B) is much faster than learning by listening (E). Neural cryptography is much simpler than the commonly used algorithms [11] [12].

*3.2. Algorithm*

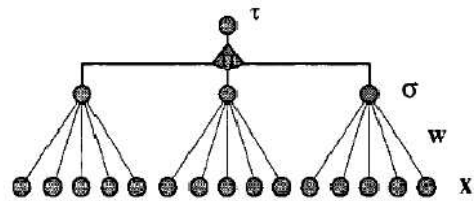

Figure 2: *Tree Parity Machine*

Here is a simple neural network as shown in figure 2. It consists of an input vector x, a hidden layer sigma, a weights coefficients w between input vector and the hidden layer which is an activation procedure that counts the result value t. Such a neural network is called as neural machine. It is described by three parameters: K-the number of hidden neurons, N-the number of input neurons connected to each hidden neuron, and L-the maximum value for weight {-L...+L}. Two partners have the same neural machines. Output value is calculated by

$$\tau = \pi \prod_{i=1}^{K} SIGN \left[ \sum_{j=1}^{N} w_{i,j}\ x_{i,j} \right]$$

We update the weights only if the output values of neural machines are equal. There are three different rules:

- Hebbian learning rule:
$$w_{i,j}^{+} = g(w_{i,j} + x_{i,j}\ \tau\theta(\sigma_i\tau)\theta(\tau^A\tau^B))$$

- Anti-Hebbian learning rule:
$$w_{i,j}^{+} = g(w_{i,j} - x_{i,j}\ \tau\theta(\sigma_i\tau)\theta(\tau^A\tau^B))$$

- Random-walk learning rule:
$$w_{i,j}^{+} = g(w_{i,j} + x_{i,j}\ \theta(\sigma_i\tau)\theta(\tau^A\tau^B))$$

## 4. SECRET KEY GENERATION

### 4.1. Key Generation

The different stages in the secret key generation procedure which is based on neural networks can be stated as follow [13]: as shown in figure 3.

1. Determination of neural network parameters: k, the number of hidden layer units n, the input layer units for each hidden layer unit l, the range of synaptic weight values is done by the two machines A and B.
2. The network weights to be initialized randomly.
3. The following steps are repeated until synchronization occurs.
4. Inputs are generated by a third party (say the key distribution centre).
5. The inputs of the hidden units are calculated.
6. The output bit is generated and exchanged between the two machines A and B.
7. If the output vectors of both the machines agree with each other then the corresponding weights are modified using the Hebbian learning rule, Anti-Hebbian learning rule and Random-walk learning rule.
8. When synchronization is finally occurred, the synaptic weights are same for both the networks. And these weights are used as secret key.

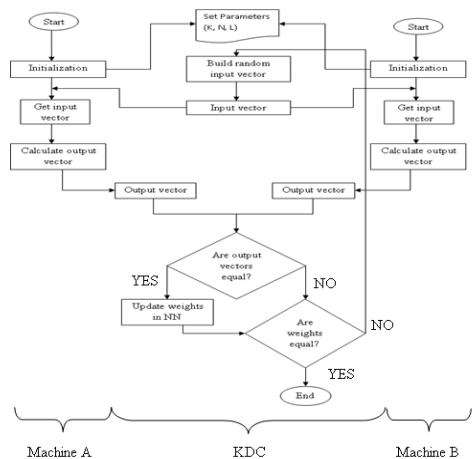

Figure 3: *Key Generation*

## 5. IMPLEMENTATION

In this section we will describe how to program neural machines and will show how to use MATLAB. The figure 3 shows how to synchronize the two machines. The main function used here is tTPM (TPMTree Parity Machine [14]). It contains vectors: H and W. 'H' is used for internal operations during result value counting. 'W' contains weights. There are also four integer values: K, L, N, and TPOutput. Here in every iteration, we should produce the input vector, the count output value by using functions tInputVector() and CountResult(). Whenever the output of the two machines is same then the weights are updated using UpdateWeight() function. The FormRandomVector() function is used to find the random input vectors by the key distribution centre. To find the random bit the randi function from MATLAB is used which uniformly distributes pseudorandom integers.

## 6. RESULT

### 6.1. Analysis on Neural Network Key Generator

Table 1: *Result table for Neural Network Key Generator*

| Sl.No | Different Issues | With NN | Without NN |
|---|---|---|---|
| 1. | Synchronization time | Required | Not required |
| 2. | Randomness | More | No |
| 3. | Security | More | Less |

### 6.1.1. Synchronization Time

The data set obtained for synchronization time by varying number of input units (n) is shown in figure 4. The number of iterations required for synchronization by varying number of input units (n) is shown in figure 5. The two figures show that as the value of n increases, the synchronization time and number of iterations also increases. The %iterations (actual/max) required for synchronization by varying number of input units (n) is shown in figure 6. The %synchronization time per iteration required for synchronization by varying number of input units (n) is shown in figure 7. In these two figures it is shown that as the value of n increases, the %iterations (actual/max) and %synchronization time per iteration is decreased.

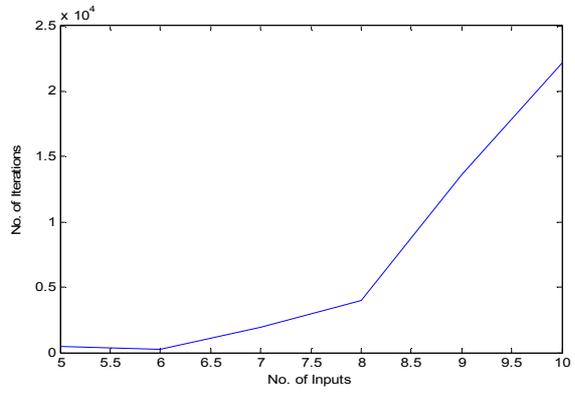

Figure 4: *No of Input Units Vs No of Iteration*

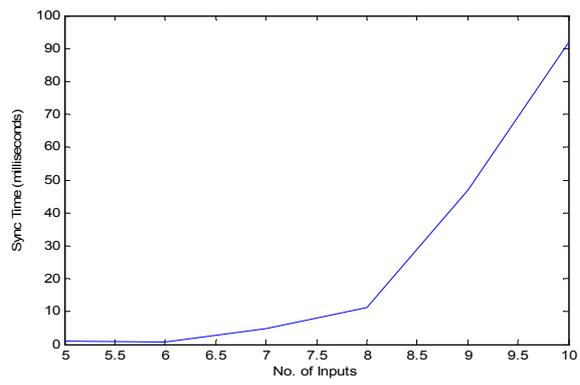

Figure 5: *No of Input Units Vs Sync. Time*

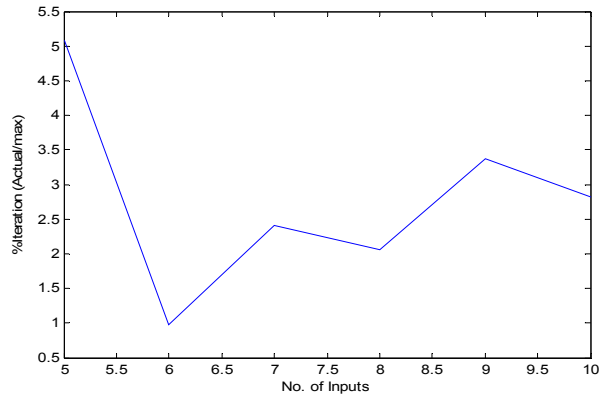

Figure 6: *No of Input Units Vs %Iteration*

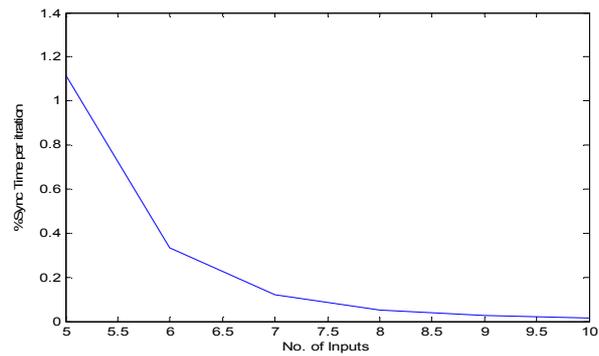

Figure 7: *No of Input Units Vs Sync. Time per Iteration*

### 6.1.2. Randomness

A random process is one whose consequences are unknown. Intuitively, this is why randomness is crucial in our work because it provides a way to create information that an adversary can't learn or predict. When speaking about randomness, we commonly mean a sequence of independent random numbers, where each number was obtained completely random and has absolutely no correlation between any other numbers inside the sequence. Here in our work in each iteration we are going to get the different keys, hence randomness is more as shown in figure 8. So here an attacker cannot predict the key.

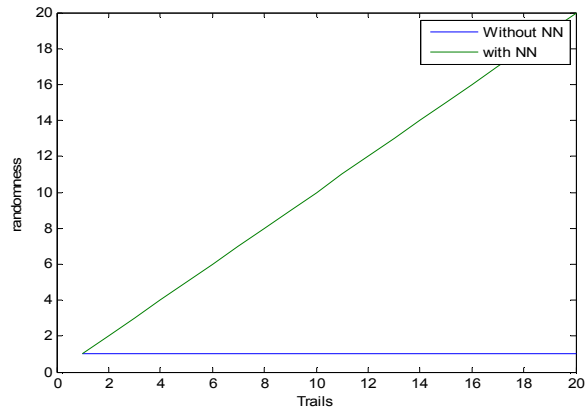

Figure 8: *Randomness Vs No. of Trials*

Here we can say that security is directly proportional to randomness, hence we have achieved security as well.

### 6.1.3. Security

Security is directly proportional to randomness; hence we have achieved security as well.

### 6.2. Analysis on Agent Approach

Table 2: *Result Table for Agent Approach*

| Sl.No | Different Issues | With Agent | Without Agent |
|---|---|---|---|
| 1. | Distribution time | Less | More |
| 2. | Attack | Not Possible | Possible |
| 3. | Alternate on attack | Yes | No |

To analyze on distribution time here the mobile agents are compared with general client server model. In client server model there will be continuous exchange of request and response messages and with mobile agents there is no continuous network communication. Figure 9 shows the graph of number of reading versus distribution time which is in milliseconds. As shown in figure 9 the distribution time by using mobile agents is less as compared with client server model.

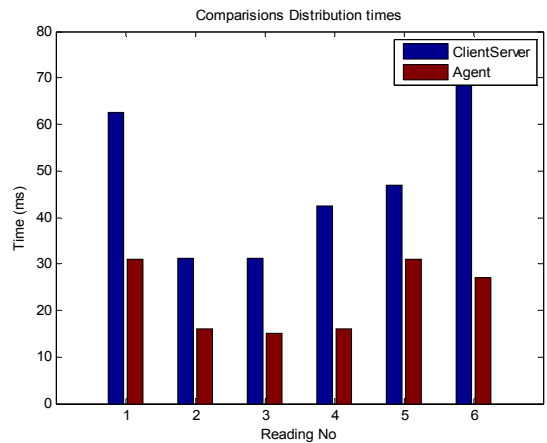

Figure 9: *Number of reading Vs Time*

7. CONCLUSION

Interacting neural networks have been calculated analytically. At each training step two networks receive a common random input vector and learn their mutual output bits. A new phenomenon has been observed: Synchronization by mutual learning. The two partners can agree on a common secret key over a public channel. An opponent who is recording the public exchange of training examples cannot obtain full information about the secrete key used for encryption .This works if the two partners use multilayer networks, parity machines. We have shown graphs by which we can come to know that synchronization time will go on decreasing as the number of inputs increase. The opponent has all the information (except the initial weight vectors) of the two partners and uses the same algorithms. Nevertheless he does not synchronize. Here we have also achieved the randomness of key. The use of multi-agent approach is motivated by the system functioning in heterogeneous environment, and by processing data in different operating systems.

## 8. FUTURE WORK

The key distribution centre generated the secret key. Our future work is that the key distribution centre will distribute the generated key securely by some method.

SAHANA S.BISALAPUR
M.TECH COMPUTER SCIENCE AND ENGINEERING
CONTACT: sahanabisalapur@gmail.com